# The long-term steady motion of Saturn's Hexagon and the stability of its enclosed jet-stream under seasonal changes

A. Sánchez-Lavega[1,2*], T. del Río-Gaztelurrutia[1-2], R. Hueso[1-2], S. Pérez-Hoyos[1-2], E. García-Melendo[1,3], A. Antuñano[1], I. Mendikoa[1], J. F. Rojas[1-2], J. Lillo[4], D. Barrado-Navascués[4], J. M. Gomez-Forrellad[3], C. Go[5,9], D. Peach[6,9], T. Barry[7,9], D. P. Milika[9], P. Nicholas[9], A. Wesley[8,9] and the IOPW-PVOL Team[9 †]

[1]Departamento de Física Aplicada I, E.T.S. Ingeniería, Universidad del País Vasco, Alameda Urquijo s/n, 48013 Bilbao. Spain.

[2]Unidad Asociada Grupo Ciencias Planetarias UPV/EHU- IAA (CSIC). Spain

[3]Fundació Observatori Esteve Duran, Montseny 46, 08553 Seva, Barcelona. Spain.

[4] Departamento de Astrofísica, Centro de Astrobiología (INTA-CSIC), Campus ESAC (ESA), P.O.Box 78, 28691 Villanueva de la Cañada. Spain.

[5]Physics Department, University of San Carlos. Address: Nasipit, Talamban, Cebu City. Philippines.

[6]British Astronomical Association, Burlington House, Piccadilly, London W1J 0DU. UK.

[7]Astronomical Society of New South Wales, Broken Hill Observatory, NSW 2880. Australia.

[8]Astronomical Society of Australia, School of Physics, The University of Sydney NSW 2006 Australia

[9]International Outer Planet Watch - Planetary Virtual Observatory. A list of participants of the International Outer Planet Watch Team IOPW-PVOL appears in the Supplementary Infromation

*To whom correspondence should be addressed. E-mail: agustin.sanchez@ehu.es

[†] Partially based on observations obtained at Centro Astronómico Hispano Alemán, Observatorio de Calar Alto MPIA-CSIC, Almería. Spain

We investigate the long-term motion of Saturn's North-Pole Hexagon and the structure of its associated eastward jet, using Cassini ISS and ground-based images from 2008 to 2014. We show that both are persistent features that have survived the long polar night, the jet profile remaining essentially unchanged. During those years the hexagon vertices showed a steady rotation period of 10 hr 39 min 23.01 ± 0.01 s. Analysis of Voyager 1 and 2 (1980-1981) and HST and ground-based (1990-91) images shows a period shorter by 3.5s, due to the presence at the time of a large anticyclone. We interpret the hexagon as a manifestation of a vertically trapped Rossby wave on the polar jet and, because of their survival and unchanged properties under the strong seasonal variations in insolation, we propose that both hexagon and jet are deep-rooted atmospheric features that could reveal the true rotation of the planet Saturn.

**1. Introduction**

Images from Voyager 1 and 2 flybys in 1980 and 1981 showed a hexagonal feature at planetocentric latitude 75°N enclosing a strong eastward jet with peak speed of 120 ms$^{-1}$ (*Godfrey*, 1988). The same feature was later re-observed in 1990-1995 with ground-based telescopes and the Hubble Space Telescope (*Sánchez-Lavega et al.*, 1993; *Caldwell et al.*, 1993). At those times, a large anticyclone ~ 11,000 km in length (the North Polar Spot, NPS) was present in the region outside the hexagon and it was proposed that the NPS impinging on the jet generated the hexagonal wave (*Allison et al.*, 1990). Because of this relationship, the NPS was assumed to move with the hexagon and its motion was put forward as a possible imprint of Saturn's rotation (*Godfrey*, 1990; *Sánchez Lavega et al.,* 1997). The North Polar Region was not re-observed until 2007-2008 when CIRS and VIMS instruments onboard the Cassini spacecraft revealed the Hexagon in Saturn's night both in the temperature field (*Fletcher et al.*, 2008) and in the cloud opacity at 5 μm (*Baines et al*., 2009). In this paper we present a study of the region during 5.5 years (2008-2014), determining the precise motion of the hexagon vertices, the absence of the NPS and the persistence and unchanging properties of the embedded eastward jet stream despite the strong seasonal radiation forcing at Saturn's poles (*Pérez-Hoyos and Sánchez-Lavega,* 2006b). In addition, we present the first analysis of the motion of the hexagon vertices in 1980-81 and 1990-91, extending the studied period to 33 years and allowing us to contextualize the movement of the hexagon in its relationship to the NPS.

**2. Image selection and measurement method**

Two main data sets have been analyzed along the period 2008-2014: (1) Cassini ISS images (*Porco et al.*, 2004) obtained from 25 August 2008 to 28 November 2012 with the CB2 filter (wavelength 752 nm) with a typical spatial resolution ranging from 22 to 170 km/pix. These images were gathered from the NASA PDS node; (2) Ground-based images captured with telescopes ranging from 0.36 to 2.2 m in aperture operating under the "lucky-imaging methodology" (e.g. Law et al., 2006) obtained from December 2012 to January 2014 in the 380 nm – 1 μm wavelength range and with a typical spatial resolution of 1900 km. A large number of these images were submitted by observers around the world to the IOPW-PVOL database (*Hueso et al.,* 2010a). The data set was complemented with images obtained in different runs along this period with the PlanetCam UPV-EHU (*Sánchez-Lavega et al.,* 2012) and AstraLux (*Hormuth et al.,* 2008) instruments mounted on the 1.23 m and 2.2 m telescopes at Calar Alto Observatory (Southern Spain) that cover the visual range (380 nm – 1 μm) at specific selected wavelengths. Figure 1 shows a set of images of the Polar Region. The hexagon stands out as a distinct albedo band outlined by outer and inner edges that also indicate the width of the eastward jet.

The Cassini images were navigated, polar projected and measured using the software *PLIA* (*Hueso et al.*, 2010b). Two different and complementary software packages, *LAIA* (*Cano*, 1998) and *WinJupos* (*Hahn and Jacquesson*, 2011) were used to navigate and measure the ground-based images. The position (longitude in System III and planetocentric latitude) of each hexagon "corner" or vertex in the outer and inner edges, were measured with a typical error of 2º in longitude but less than 0.5º in latitude.
The wind profile of the eastward polar jet where the hexagon resides was measured on Cassini ISS CB2 images (752 nm) corresponding to three periods: 29 November to 1 December 2008, 2-21 January 2009, and 28 November 2012. Three methods were employed to retrieve the wind profile: cloud tracking on image pairs separated by one planetary rotation (about 10 hr), brightness correlation of one-dimensional zonal scans along each hexagon side at different latitudes (*García-Melendo et al.*, 2011), and two-dimensional brightness correlation over polar projected images (*Hueso et al*, 2009). Typical velocity errors coming from navigation and tracer misidentifications are ~ 10-15 ms$^{-1}$.

In addition, we include new measurements of a set of Voyager 1 and 2 images in 1980 and 1981, and of a few available Hubble Space Telescope and ground-based images in 1990-91 that display the feature. Details of the method of analysis of these images can be found in *Sánchez-Lavega et al.* (1993, 1997, 2000). The list of analyzed images are given in the Supplementary Information.

## 3. Results

The 2008-2014 analysis shows a mean planetocentric latitude of the six vertices of the hexagon of 74.7º ± 0.2º, with a mean North-South width of 2.8º ± 0.5º. A linear fit to the longitude drift rates for each of the six vertices (Table 1, Figure 2A) results, when combined, in a mean drift rate for the hexagon of ω = +0.0128 ± 0.0013°/day relative to System III rotating frame, with angular velocity $\Omega_{III}$ = 810.7939024 °/day (*Desch and Kaiser*, 1981; *Seidelmann et al.*, 2007). We also used the LBG vector quantization method (*Linde et al.*, 1980; *Gray*, 1984) to determine the hexagon's rotation period. In our case vector values are the fixed position in longitude and latitude of the hexagon's six vertices. We look for the drift rate that, once subtracted, simultaneously minimizes the dispersion of our position measurements about six optimally chosen vertex positions. To automatically compute vertex positions we use Lloyd's iterative technique (*Lloyd*, 1957). This method gives ω = +0.0131 ± 0.0010 deg day$^{-1}$ (Figure 2B), in excellent agreement with the previous one. The mean drift rate combining both methods is <ω> = +0.0129 ± 0.0020 deg day$^{-1}$. This yields an absolute angular velocity for the hexagon <Ω> = 810.7810 °/day that translates into a hexagon speed relative to System III of -0.036 ms$^{-1}$. The corresponding absolute rotation period for the hexagon in 2008-2014 is <τ> = 10 hr 39 min 23.01 ± 0.01 s.

A track of the motion of the hexagon vertices in 1980-81 shows also a small drift but in opposite direction with <ω> = - 0.0602 ± 0.014 deg day$^{-1}$ (<τ> = 10 hr 39 min 19.6 ± 0.7 s), close to that of the NPS at the time ($<\omega>$ = - 0.0444 ± 0.01 deg day$^{-1}$; $<\tau>$ 10 hr 39 min 20.3 ± 0.5 s) (Table 1 and Figure 2C; *Godfrey* 1988; *Sánchez-Lavega et al.*, 1993, 1997). The small number of available data for 1990-91 shows a lower drift rate for the hexagon (Figure 2D; $<\omega>$ = - 0.0010 ± 0.007 deg day$^{-1}$), again similar to that of the NPS, indicating that the motion of both features (hexagon and NPS) was coupled. The NPS anticyclone, initially proposed to originate the hexagon (*Allison et al.*, 1990), is no longer present in the images from 2008-2014 and it was therefore a transient vortex. We conclude that the NPS is not necessary for the existence of the hexagon, but when present, it could have slightly disturbed the hexagon motion.

The hexagon encloses an intense and narrow eastward jet. Its velocity profile was measured in 1980-81 at the altitude level corresponding to a pressure 0.5–1 bar (*Godfrey*, 1988) and again in 2008 at level ~ 2-4 bar (*Baines* et al., 2009). Here we present new measurements of the jet in 2008, 2009 and 2012 based on Cassini ISS CB2 images that approximately sense the same altitude level as the Voyager images (*García-Melendo et al.*, 2011). Figure 3 shows the wind profile in 2008-09 and 2012 and compares it with that of 1980-81 from the Voyagers. It is evident that there have been no significant changes on the jet intensity and shape, in spite of the marked seasonal insolation variability at those latitudes (*Pérez-Hoyos and Sánchez-Lavega*, 2006b). Geostrophic conditions prevail for this jet since the Rossby number $Ro = u/f_0 L \sim 0.13$, with a Coriolis parameter $f_0 = 2\Omega_{III} \sin\varphi_0 = 3.17 \times 10^{-4}$ s$^{-1}$, a jet peak velocity of $u = 120$ ms$^{-1}$ at planetocentric latitude $\varphi_0 = 75.5°$ N and $L = 2,900$ km the jet width at half maximum. According to our data, the jet has a peak averaged curvature $-<d^2u/dy^2> \approx 6\text{-}8 \times 10^{-11}$ m$^{-1}$s$^{-1}$ (Figure 3) or $-<d^2u/dy^2>/\beta \approx 20-40$ violating the Rayleigh-Kuo instability criterion for a purely zonal jet both in a shallow barotropic atmosphere (*Sánchez-Lavega*, 2011) or in a deep atmosphere (*Ingersoll and Pollard*, 1982). Here $y$ is the meridional coordinate, $\beta = (df/dy) = 2\Omega_{III}\cos\varphi_0/R_S = 1.5 \times 10^{-12}$ m$^{-1}$s$^{-1}$ is the planetary vorticity gradient and $R_S = 55,250$ km is the radius of Saturn at the jet latitude.

**4. Interpretation**

An early Rossby-wave model for the hexagon (*Allison et al.*, 1990) simplified the atmospheric structure and proposed the wave to be forced by the NPS, which is not currently present. Two other studies have been presented to explain the origin of the hexagon: a barotropic linear instability analysis together with laboratory experiments of a fluid in a rotating tank (*Barbosa-Aguiar et al.*, 2010) and a non-linear numerical model using the EPIC code (*Morales-Juberías et al.*, 2011). Both studies predict a conspicuous coherent vortex street pattern that is not observed in Cassini images (Figure 1) and numerical simulations using the EPIC model (*Morales-Juberías et al.*, 2011) predict a non-consistent phase speed for the hexagon pattern.

Here we interpret the hexagon as a stationary trapped Rossby wave that could be the manifestation of the meanders of a deep unstable polar jet. The three-dimensional Rossby wave is represented by a stream function $\psi'(x,y,z,t) = \psi_0 \exp\left[(z/2H) + i(kx + ly + mz - \omega_0 t)\right]$ that for quasigeostrophic conditions in a stratified atmosphere with no vertical shear ($\partial u/\partial z = 0$) obeys the dispersion relationship (*Vallis*, 2006; *Sánchez-Lavega*, 2011)

$$c_x - u = -\frac{\beta - <d^2u/dy^2>}{k^2 + l^2 + \left(\frac{f_0^2}{N^2}m^2\right) + \frac{1}{4L_D^2}} \quad (1)$$

Here $\psi_0$ is the wave amplitude, $k = 2\pi/L_x, l = 2\pi/L_y, m = 2\pi/L_z$ the zonal, meridional and vertical wavenumbers respectively, $L_D=NH/f_0 \sim 1,000$ km the Rossby deformation radius at troposphere (*Read et al.*, 2009), $\omega_0$ the wave frequency, $H=(R_g^*T)/g$ the atmospheric scale-height (*T* the temperature, *g* the gravity acceleration, $R_g^* = 3892$ Jkg$^{-1}$K$^{-1}$ the gas constant) and *N* the Brunt-Väisälä frequency. The zonal and meridional wavenumbers $k = 2\pi/L_x$ and $l = 2\pi/L_y$ are calculated taking $L_x=14,500$ km from the hexagon zonal wavenumber 6 (hexagon side length), and $L_y=5,785$ km that assumes that the hexagon meridional extent (3º in latitude) corresponds to half the meridional wavelength. We take the hexagon velocity in System III in 2008-2014 as its wave speed $c_x = -0.036$ ms$^{-1}$. From the jet profile (Figure 3) we take $u = 120$ ms$^{-1}$ ($c_x << u$) yielding $-<d^2u/dy^2>/u \approx 6\times10^{-13}$ m$^{-2}$. From these numbers we see in (1) that $\ell^2 > \left[k^2 + (0.25L_D^{-2})\right]$ and $\ell^2 > \left[<d^2u/dy^2>/(-u)\right]$. Therefore, the zonal wavenumber dominates because of the strong eastward flow velocity. This implies that $m^2 < 0$, and the Rossby wave is vertically trapped within the region of positive static stability in Saturn's atmosphere in agreement with a previous analysis (*Allison et al.,* 1990). The vertical extent of this trapped region for the wave is unknown and can penetrate well below the lower water clouds at ~ 10 bar altitude level. To have $m^2 > 0$ (i.e. for vertical propagation with real $L_z$) the above numbers require $L_y > 15,700$ km which is highly unrealistic since the meridional wave oscillation will reach the pole in the north and overcome the neighboring jet in the south.

**5. Discussion**

Because of Saturn's rotation axis tilt of 26.7º, the planet is subject to strong seasonal insolation changes along its year of 29.475 Earth's years (*Pérez-Hoyos and Sánchez-Lavega*, 2006b), with long alternating periods of light and darkness that affect temperatures above the ~ 2 bar altitude level (*Bézard et al.,* 1984; *Sánchez Lavega et al.,* 1997; *Fletcher et al.*, 2010). The survival of the hexagon and the stability of the polar jet to the varying radiative forcing suggest that they extend deep in the atmosphere. This is in agreement with what has been deduced for other Saturn jets from studies of a long-lived anticyclone (*García-Melendo et al.*, 2007) and from the giant Great White Spot storm in 2010 (*Sánchez-Lavega et al.,* 2011; *García-Melendo et al.*, 2013). From those works, the zonal winds were found to extend down to the ~ 10 bar level or deeper with no vertical shear.

Currently there is an open debate on Saturn's rotation period (*Sánchez-Lavega A.*, 2005; *Gurnett et al.,* 2010). The rotation is determined by the periodic radio emissions with origin in the magnetic field, assumed to be tied to the deep interior. However, Saturn's rotation period is not yet determined since the radio measurements from Saturn's Kilometric Radiation (SKR) have varied in the last 30 years showing two simultaneous periods that differ by 15 min, one for each hemisphere (*Gurnett et al.,* 2010). Alternatives based on Saturn's gravitational field (*Anderson et al.,* 2007) and on the stability of the atmospheric zonal jet stream pattern (*Read et al.,* 2009) also differ from

6each other and are 5 to 7 minutes shorter than the mean SKR period. Our long-term study shows that the hexagon has indeed an extremely steady rotation period, with its associated jet remaining unchanged by the strong seasonal cycle, properties that are consistent with deeply rooted features. We suggest that the current steady rotation period of the hexagon of 10 hr 39 min 23.01 s ± 0.01 s, free of disturbances from the NPS, could represent the period of Saturn's internal "solid body" rotation.

**Acknowledgements** We gratefully acknowledge the work of the Cassini ISS team that made the data available. This work was supported by the Spanish MICIIN projects AYA2009-10701, AYA2012-38897-C02-01, AYA2012-36666 with FEDER support, PRICIT-S2009/ESP-1496,Grupos Gobierno Vasco IT765-13 and UPV/EHU UFI11/55.**References**

Anderson, J. D. and G. Schubert (2007), Saturn's Gravitational Field, Internal Rotation, and Internal Structure, *Science*, 317, 1384-1387.

Allison M., D. A. Godfrey and R. F. Beebe (1990), A wave dynamical interpretation of Saturn's Polar Hexagon, *Science*, 247, 1061-1063.

Baines K. H. et al. (2009), Saturn's north polar cyclone and hexagon at depth revealed by Cassini/VIMS, *Planet. Space Sci.*, 57, 1671-1681.

Barbosa-Aguiar A. C. et al. (2010), A laboratory model of Saturn's North Polar Hexagon, *Icarus*, 206, 755–763.

Bézard B., D. Gautier, B. Conrath (1984), A seasonal model of the Saturnian upper troposphere: Comparison with Voyager infrared measurements. *Icarus*, 60, 274-288.

Caldwell J. et al. (1993), An observed drift of Saturn's Polar Spot by HST, *Science*, 260, 326-329.

Cano J. A., *L.A.I.A.: Laboratorio de Análisis de Imágenes Astronómicas* (Grup d'Estudis Astronòmics, Barcelona, 1998).
Desch M. D., L. M. Kaiser (1981) Voyager measurements of the rotation period of Saturn's magnetic field, *Geophys. Res. Lett.*, 8, 253-256.

Fletcher L. N. et al. (2008), Temperature and Composition of Saturn's Polar Hot Spots and Hexagon, *Science*, 319, 79-81.

Fletcher L. N. et al. (2010), Seasonal change on Saturn from Cassini/CIRS observations, 2004–2009, *Icarus*, 208, 337 – 352.

García-Melendo E., A. Sánchez-Lavega, R. Hueso (2007), Numerical models of Saturn's long-lived anticyclones, *Icarus*, 191, 665-677.

García-Melendo E., et al., (2011), Saturn's zonal wind profile in 2004 - 2009 from Cassini ISS images and its long-term variability, *Icarus,* 215, 62-74.

**Table 1: Individual drift rates of the Hexagon vertices in 2008-2013 and in 1980-81**

| Drift rate (°/day) (2008-2014) | Mean Latitude (2008-2014) | Drift rate (°/day) (1980-81) |
|---|---|---|
| 0.0107± 0.0005 | 74.5± 1.4 | -0.0757 |
| 0.0125± 0.0003 | 74.6± 1.1 | -0.0594 |
| 0.0133± 0.0003 | 74.8± 1.3 | -0.0787 |
| 0.0130± 0.0005 | 75.0± 1.2 | -0.045 |
| 0.0126± 0.0004 | 74.8± 1.0 | -0.052 |
| 0.0146± 0.0005 | 74.6± 1.5 | -0.0505 |
| Mean value | Mean value | Mean value |
| 0.0128 ± 0.0013 | 74.7± 0.2 | -0.0602± 0.0014 |

**Note:** The drift rate ($\omega$) is relative to System III Longitude and is the slope of the linear fits shown in Figure 2. Latitudes are given in planetocentric degrees. The error of the mean values is calculated as the standard deviation of the average values for individual vertices.

**FIGURE CAPTIONS**

**Figure 1**. Polar projected images of Saturn from the North Pole to 60ºN. (A) Composition of Cassini ISS CB2 images on January 3, 2009; (B) Cassini ISS RGB color composition showing features in the hexagon and jet stream on August 26, 2008. The red arrow marks the location of the hexagon; (C) Color composite from July 13-16, 2013 obtained with the AstraLux camera on the 2.2 m telescope at Calar Alto Observatory (Spain); (D, E, F) Images obtained in 2013 by D. P. Milika (March 25), D. Peach (April 20) and C. Go (May 26), respectively.

**Figure 2**. Motion of the Hexagon vertices. (A) Longitude drifts in System III of the six vertices between August 2008 and January 2014. The straight lines represent linear fits to each vertex drift in longitude; (B) Dispersion diagram for a drift range between -10.0 to +10.0 deg day$^{-1}$ with respect to System III. There is a distinct minimum that marks the rotation period. (C) Longitude drifts in System III of the six vertices and NPS (red squares) between October 1980 and August 1981 (Voyager 1 and 2 images); (D) Longitude drifts in System III of the six vertices and NPS (red squares) between July and November 1990 (HST and Pic-du-Midi images

**Figure 3.** Measurements of the hexagon polar jet in 2008-09 and 2012 using Cassini ISS images. (A) Zonal wind profile: dark continuous line (1D correlation, 29 Nov. – 1 Dec. 2008), crosses (cloud tracking, 2-3 January 2009), circles (2D correlation, 28 Nov. 2012). A comparison is performed with Voyager 1 and 2 measurements in 1980-1981 (gray line and small dots) (*Godfrey,* 1988). The red parabolic curve indicates the jet peak intensity from individual cloud tracking; (B) Ambient vorticity profiles smoothed for two degrees resolution (dark and grey lines) with the variable part of the planetary vorticity ($\beta y$) indicated by the red line; (C) Profile of the jet curvature with the same smoothing (dark and grey lines). The planetary vorticity gradient ($\beta$) is shown as a red line.

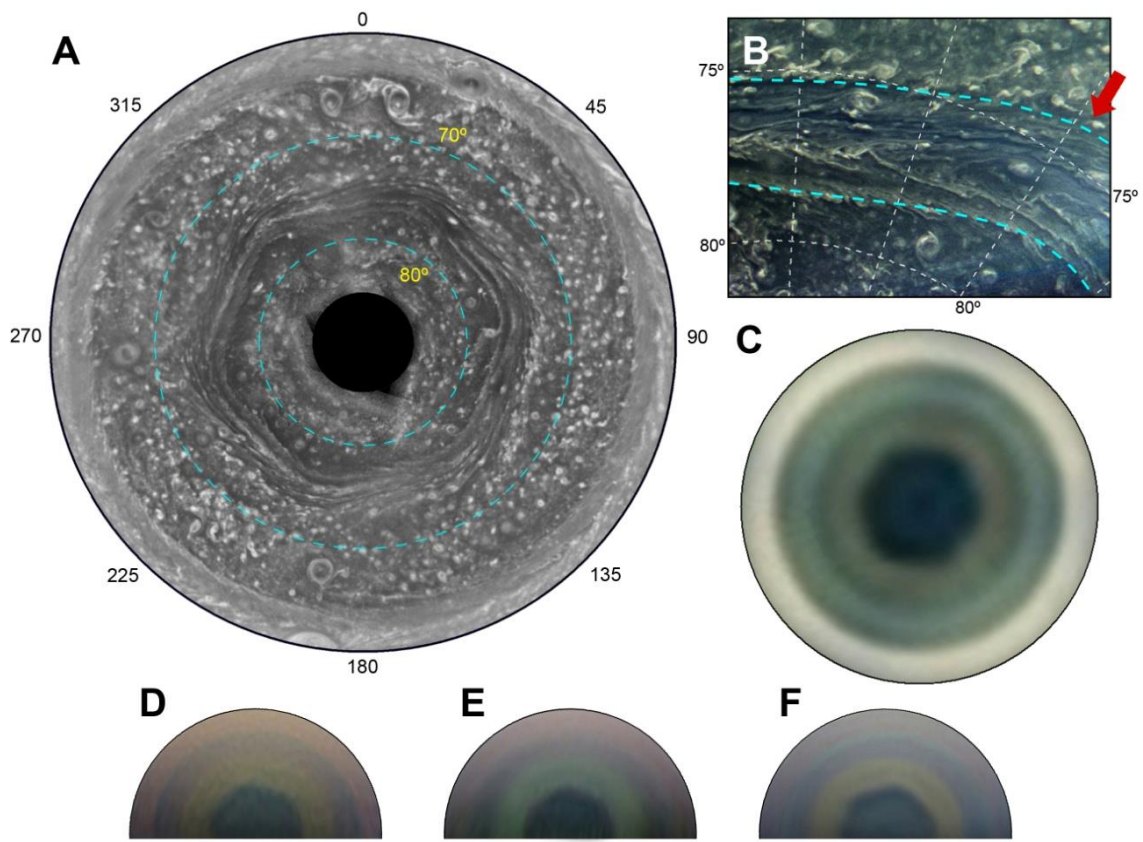

**FIGURE 1**

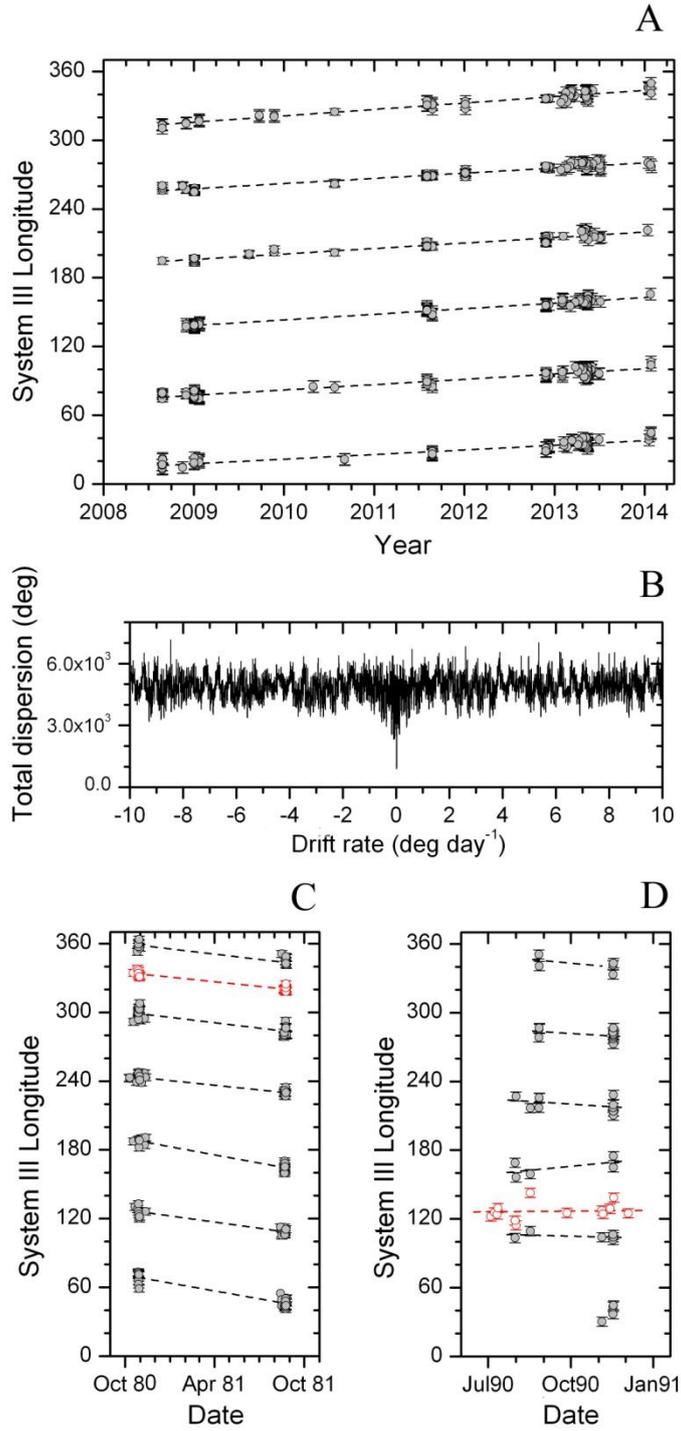

**FIGURE 2**

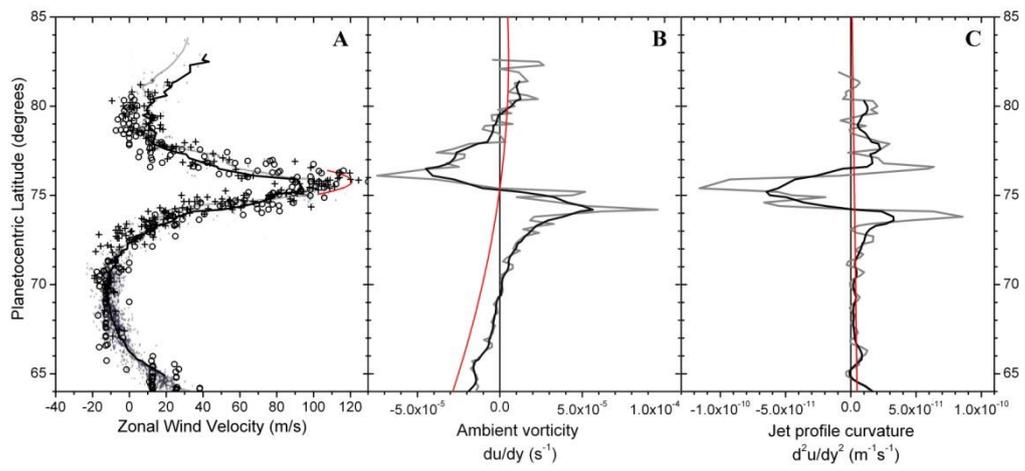

**FIGURE 3**